# Neurofeedback: principles, appraisal and outstanding issues

David Papo

*SCALab, CNRS, Université Lille3, Villeneuve d'Ascq, France*
*E-mail address: papodav@gmail.com*

___________________________________________________________________________________

ABSTRACT

___________________________________________________________________

Neurofeedback is a form of brain training in which subjects are fed back information about some measure of their brain activity which they are instructed to modify in a way thought to be functionally advantageous. Over the last twenty years, NF has been used to treat various neurological and psychiatric conditions, and to improve cognitive function in various contexts. However, despite its growing popularity, each of NF's main steps comes with its own set of often covert assumptions. Here we critically examine some conceptual and methodological issues associated with the way NF's general objectives and neural targets are defined, and review the neural mechanisms through which NF may act, and the way its efficacy is gauged. The NF process is characterised in terms of functional dynamics, and possible ways in which it may be controlled are discussed. Finally, it is proposed that improving NF will require better understanding of various fundamental aspects of brain dynamics and a more precise definition of functional brain activity and brain-behaviour relationships.

___________________________________________________________________________________

## 1. Introduction

As knowledge of brain structure and dynamics dramatically progresses, neuroscientists approach a stage wherein the brain can not only be described but also acted upon in increasingly controlled, and even constructive and enhancing fashions (Deca and Koene, 2014; Sitaram et al., 2017; Bassett and Sporns, 2017; Medaglia et al., 2017).

Of the many brain intervention strategies, including brain-computer interfaces (BCI), deep brain stimulation (DBS), transcranial magnetic stimulation (TMS), or optogenetics, neurofeedback (NF) is possibly the most conceptually intriguing. NF trains subjects to self-regulate a measure extracted in real time from their own brain activity, typically recorded with non-invasive devices (Coben and Evans, 2011; Marzbani et al., 2016). This measure is somehow associated with performance of a cognitive, motor or neurophysiological function. Remarkably, people appear to be able to steer some aspects of their own behaviour by virtue of information on how their own brain is handling it, thereby shedding light on the way the power of thought, as it were, normally promotes behavioural dynamics.

NF has been used in a range of cognitive (Gruzelier, 2014a), psychiatric (Fovet et al., 2015; Marzbani et al., 2016) and neurological conditions (Marzbani et al., 2016), including emotion regulation (Johnston et al., 2010; Zotev et al., 2013; Keinan et al., 2016; Jacob et al., 2018), attention-deficit hyperactivity disorder (ADHD) (Gevensleben et al., 2012; Sonuga-Barke et al., 2013; Vollebregt et al., 2014a; Micoulaud-Franchi et al., 2014a; Schönenberg et al., 2017), autism (Coben et al., 2010; Pineda et al., 2014; Friedrich et al., 2015; LaMarca et al., 2018), depression (Linden et al., 2012; Young et al., 2014; Zotev et al., 2016), schizophrenia (Surmeli et al., 2012; Dyck et al., 2016), addictions (Arani et al., 2010; Unterrainer et al., 2013; Gerchen et al., 2018), eating disorders (Lackner et al., 2016), post-traumatic stress disorder (Kluetsch et al., 2014), epilepsy (Tan et al., 2009; Sterman, 2010; Micoulaud-Franchi et al., 2014c), stroke (Rozelle and Budzynski, 2001; Mihara et al., 2013; Wang et al., 2017), traumatic brain injury (Thornton and Carmody, 2008; May et al., 2013; Munivenkatappa et al., 2014; Bennett et al., 2017), pain (deCharms et al., 2005), and insomnia (Cortoos et al., 2010; Schabus et al., 2017). Furthermore, NF protocols have used non-invasive recording techniques including electroencephalography (EEG), magnetoencephalography (MEG), functional magnetic resonance imaging (fMRI) and near infra-red spectroscopy (NIRS) (Weisskopf et al., 2004; Foldes et al., 2015; Okazaki et al., 2015; Keynan et al., 2016; Sepulveda et al., 2016; Marzbani et al., 2016; Liu et al., 2017).

The range of aspects of brain activity that *can* be modulated in NF procedures seems quite large, but which ones *should* be modulated is still a matter of debate. The remainder of this article addresses the following outstanding questions: how are NF's general objectives and neural targets defined? On what neural mechanisms does NF act? How is its efficacy gauged? How and to what extent can it be improved?

## 2. Neurofeedback: conceptual underpinnings and modus operandi

NF can be thought of as a non-invasive brain stimulation technique equipped with a closed-loop control mechanism, whereby information on the dynamics, usually *non-observable*, is made observable to subjects, who can then use it to retroact on it, and push it towards *functionally desirable* goal states. NF involves defining: *i)* the general *goal*; *ii)* a neural *target* as feature; *iii)* an appropriate stimulation schedule. Each of these steps comes with its own set of (often covert) assumptions and bears profound conceptual implications for our understanding of fundamental mechanisms of brain function and dysfunction.

### 2.1. Identifying goals: task-induced vs. resting state-based neurofeedback

What goals can we hope to achieve with NF? A flavour of the range of achievable goals is given by noting that NF is in a sense a neuromimetic process. In fact, closed *feedback loops* have long been known to constitute a general principle of brain functioning (Aitchinson and Lengyel, 2017). For instance, in motor control, a forward internal model is thought to be used to generate the predicted sensory feedback that estimates the sensory consequences of a motor command. These are then compared with the corollary discharge to inform the central nervous system about how well the expected action matched its actual external one (Wolpert et al., 1995). At the short time-scales typical of sensory-motor tasks, an obvious goal for NF protocols is then task-specific optimization. This requires real-time decoding and mapping of brain activation to behaviour and can be achieved both for subjects with pathological conditions impairing normal performance, by resorting to cognitive tasks a given participant or population is known to fare poorly at, and conceivably, for healthy subjects, in which case NF would be used as a brain enhancement strategy.

But how does NF work in contexts qualitatively different from stimulus discrimination, e.g. in NF for ADHD? Here, the required intervention's goal does not aim at optimizing performance at a specific task, but at normalizing an entire repertoire of responses or symptoms, in away akin to a typical pharmacological intervention. In this case, NF should nudge the dynamics away from pathological dynamics and towards a regime characterized by the generic properties of the healthy brain. Abnormalities, in turn, may be characterized in terms of deviations from the generic properties of task-independent brain activity (Papo, 2014). This naturally leads to addressing the issue of defining these properties and ultimately the neurophysiological function of resting brain activity (Papo, 2013a, 2014).

### 2.2. Acting on brain dynamics: what, where, how

*What* aspect of brain activity and *where* in the brain to act upon represent the next steps in the NF process.

NF typically resorts to dynamical features of neural activity as targets, in the absence of a complete model for it. For instance, NF for ADHD generally attempts to increase the production of beta waves and decrease the number of slower brain waves (Monastra et al., 2005). This is perfectly justified in regard of the fact that neuronal network rhythmic activity at specific frequency bands is thought to contribute to information transfer and processing in the brain (Engel et al., 2001; Fries, 2005). In fMRI-guided NF protocols, the target function is often the BOLD signal amplitude in well-defined brain regions (Zotev et al., 2011, 2018; Young et al., 2014; Caria and de Falco, 2015; Hellrung et al., 2018). However, while the relevant target features is in general characterized as a scalar in an appropriate space, it could in principle be any function of possibly spatially (Ruiz et al., 2014; Koush et al., 2017) and temporally (Diaz Hernandez et al., 2016) non-local dynamics. More generally, features are *de facto* thought of as *control parameters* for the behaviour, which should in principle map into a trajectory from a given current state to a target one. An important (and largely unanswered) question is then: do commonly used targets represent genuine control parameters?

It is important to note that NF targets implicitly reflect models of brain function and the way cognitive demands and, somehow equivalently, neurological or psychiatric conditions, are understood to act on brain activity. For instance, traditional targets reflect a view whereby cognition would act as a frequency or amplitude modulator of local neuronal activity. They are also consistent with a conceptualization of neurological or psychiatric disorders as resulting from a dysfunctional activity pattern in a defined neural network that can be normalized by targeted stimulation (Christen and Müller, 2017; Braun et al., 2018). The general tenet is that stimulating/activating the (set of) regions usually responsible for the correct execution of a given cognitive task, or inhibiting those abnormally active in a given pathology should restore function/healthy behaviour. The underlying intuition is that this system is either under or over-activated due to some pathological condition, i.e. that there is a simple map between healthy and pathological dynamics in the control parameter space. However, rather than amplitude or frequency modulators, cognitive demands may be better conceived of as acting upon the functional form of dynamic brain fields (Papo, 2014). Moreover, although what is directly observable is typically dynamics, the NF target may in principle be of an altogether different form (information content; thermodynamic function, etc.).

When choosing neural targets for neuro-intervention, it seems natural to seek the anatomical region on which it should exert its effects. Anatomical localization is of course a central issue for neuro-stimulation techniques such as DBS, which requires a surgical procedure wherein an apparatus is implanted in well-specified anatomical regions. Note that, unlike in DBS or TMS, for which the target is anatomically local (though possibly distributed), in NF, the region impacted by the direct neural target to which information is delivered (e.g. primary visual or auditory areas, as feedback cues can be as simple as an audio beep or as complex as a video game) may be different from the anatomical region to be modulated by it (e.g. the amygdala).

The overall effect of acting on local targets in the anatomical space is in general highly non trivial and difficult to predict. On the one hand, brain connectivity is essential to healthy brain function and acts as a control parameter of brain dynamics (Osorio et al., 2010; Kozma and Puljic, 2015). Conversely, *dysconnectivity*, i.e. both reduced and increased connectivity, is associated with several pathological conditions (Friston, 1995; Stephan et al., 2009; Schmidt et al., 2013; Hahamy et al., 2015; Hilary and Grafman, 2017). Furthermore, both anatomical and dynamical brain connectivity can be thought of as complex networks, with non-trivial topological properties at all scales (Bullmore and Sporns, 2009). A fundamental property of networks is that perturbations to one node can affect other parts of the network, potentially causing the entire system to change its structure, dynamics and ultimately its behaviour. Thus, the consequences of targeting a spatially local region can be, and indeed generally are, complex and spatially *non-local* and multiscale in the anatomical space (Haller et al., 2013; Ros et al., 2013; Ghaziri et al., 2013; Gruzelier, 2014b; Emmert et al., 2016). For instance, striatum and anterior insula were found to be consistently activated during NF learning without being the target regions (Emmert et al., 2016). In fact, localization is a rather more general issue than that of anatomical localization - for instance non-locality in frequency has been reported (Ros et al., 2013) - and the determination of a neural target is best conceptualized as a localization process in some space, e.g. anatomical, time, frequency, or phase space.

Once the appropriate target subspace is characterized (i.e. *where* to act), brain intervention strategies must specify an action schedule (i.e. *how* to act on the chosen target). One important aspect of any stimulation schedule is represented by target activity's time scales (Papo, 2013b), as both goals and stimulation schedules may be scale-dependent. The key question in this context is: at what scales do we want stimulation to act and succeed?

At short time scales, task-specific performance optimization can be achieved by training a statistical model to discriminate brain activation patterns in response to different stimuli, decoding task-induced modifications of this pattern, adjust the stimulus based on the new brain state, and have the subject repeat the mental operation. Task complexity can eventually be altered with a time lag as short as the characteristic time scale of the target pattern, based on the distance between the current state and this pattern (Bassett et al., 2017). This goal is in principle achievable, at least in its simplest version, via a real time read-out of brain activity, but supposes temporal (and often spatial) locality of functional brain dynamics, an approximation that may often be acceptable at the short time scales of stimulus-related activity.

However, the target feature may not be approximately the same scale as the feedback loop and instead span several more orders of magnitude. This may occur when the goal is to reinstate healthy task-independent brain activity. In this case, neuro-intervention strategies should use generic properties of brain activity as guiding principles and goals. At long time scales, brain activity shows generic glassy properties, including anomalous scaling, long-range temporal correlations, weak ergodicity breaking and ageing (Bianco et al., 2007; Allegrini et al., 2011; Papo, 2014), suggesting that intervention may be temporally non-local, and other (global) properties need to be used as targets.

The need to take into account brain activity's spatial and temporal multiscale character is illustrated by DBS for Parkinson's disease. In spite of its undeniable success, DBS has so far failed to restore in patients the normal dynamical repertoire characteristic of healthy behaviour. Arguably among the reasons for the limits of this neurostimulation strategy are, on the one hand, the spatially local nature of the stimulation of a complex network, the overall outcome of which is hard to fine-tune and, on the other hand, the relative inability of current stimulation schedules to emulate the dynamical regimes induced by dopamine's phasic and tonic activity. Not only is the stimulation an open as opposed to a closed loop, but its temporal scale range is also rather narrow.

### 2.2.1. Steering dynamics

Particularly when the target feature's characteristic time is much longer than the closed-loop's average duration, it is straightforward to think of NF as a classical control problem, wherein some aspect of brain dynamics lies away from a desired (or optimal) trajectory or regime and one needs to figure out how to nudge the system so as to close in on the target dynamical trajectory or attractor. Insofar as the NF problem consists in controlling the collective dynamics of coupled nonlinearly units, control and graph theory can be used to understand how brain dynamics can be steered in a functionally advantageous way (Liu and Barabási, 2016). Within this context, it is straightforward to address the following questions: what dynamical states are accessible? Can a given target state or regime be achieved in a stable way? What's the minimal number of nodes (or links) that need to be perturbed in order to reach a given goal dynamics? What dynamical states are attainable, starting from a given initial condition?

To start addressing these questions, it is first necessary to evaluate whether brain activity is *observable* i.e. whether its dynamics can be reconstructed by monitoring its time-dependent output (Kálmán, 1963). Control could then be achieved by applying small perturbations to the system. This could be achieved by resorting to two heuristic methods, which involve altering either the system's dynamical equations (Ott et al., 1990;

Lai, 2014; Boccaletti et al., 2000) or its initial condition (Cornelius et al., 2013) to nudge the state into basin of attraction of the desired final state or attractor. These methods could in principle be used for network reprogramming and rescue from crises, e.g. with epileptic dynamics (Cornelius et al., 2013), or to engineer a particular behaviour of the system (Gutiérrez et al., 2012) by steering the dynamics towards a trajectory compatible with the system's natural dynamics but originating from a different initial condition, a procedure called *targeting* (Shinbrot et al., 1990). However, these methods generally require extensive knowledge of the system's state space, and sometimes of its dynamics (Cornelius et al., 2013), an information generally unavailable for system-level brain activity. Imperfect knowledge may, instead, cause the control strategy to push the system to the basin of attraction of an undesirable dynamics. A more feasible alternative, *target* observability, aims at identifying the sensors needed to infer the state of the system (Liu and Barabási, 2016). Importantly, the optimal sensors for network state reconstruction may not always coincide with the NF targets chosen on theoretical grounds. Furthermore, given the spatial extension, heterogeneity and inherently multiscale character of brain dynamics, it is highly non-trivial to determine both the overall effects of anatomically local stimulation and the level of coarse-graining which would guarantee good enough control targets. One should also ascertain whether the system is *controllable*, i.e. whether it can be driven from any initial condition to any desired state in finite time (Kálmán, 1963) or, more realistically, study the system's *accessibility*, i.e. the possibility to reach an open subset of the state space from a given initial state.

These important theoretical results may *prima facie* seem to be applicable to the control of neural activity. Indeed, the theoretical network control framework has started been used in neuroscience (see Bassett et al., 2017 for a review). However, these studies make rather unrealistic hypotheses on brain activity (Tu et al., 2018): brain resting dynamics is described in terms of a set of differential equations linearized around a dominant fixed point, an assumption that may account for only limited portions of the phase space, and connectivity dynamics is assumed to be linear time-invariant (Kim et al., 2017). In fact, numerical control has been shown to fail in practice even for linear systems (Sun and Motter, 2013): control trajectories turn out to be nonlocal in the phase space, i.e. the length of the state trajectory is on average independent of the distance between initial and final conditions, and there is a non-locality trade-off whereby either the control trajectory is nonlocal in the phase space or the control inputs are nonlocal in the network (Sun and Motter, 2013). Moreover, the length of the state trajectory is strongly anti-correlated with the numerical success rate and number of control inputs so that numerical control typically fails below a critical number of control inputs, (Sun and Motter, 2013). Possible solutions to this seemingly fundamental limitation have been proposed, but only for the linearized system (Klickstein et al., 2017).

On the other hand, the control of complex networks with nonlinear dynamics, and particularly of adaptive networks, is a field still in its infancy. Observability and controllability tests of such system are highly non-trivial. For adaptive systems such as the brain, in which network topology and nodal dynamics are interacting dynamical systems, and the order parameter used to describe the system's collective behaviour, may feedback onto the control parameter, limitations on network structure or dynamics may drastically constrain the controllability of the whole system.

Finally, a crucial aspect in the control of a networked dynamical system is given by its *cost*: how much "resistance" is one facing? And, in the case of a device, how much power is it going to be needed in order to achieve control? Is the required power compatible with safety? While network control methods aim at controlling dynamics through a minimal number of nodes or links, too exiguous a number may exact an excessive energetic cost (Yan et al., 2015). This issue is especially relevant for DBS devices, but it also indirectly applies to NF, and may have important implications for the determination of neural targets and, more specifically, for their anatomical locality. Remarkably, contrary to previous reports suggesting that brain resting activity may be controllable through a single node representing a given brain region (Gu et al., 2015), a recent study (Tu et al., 2018) suggested that even though brain networks might theoretically be structurally controllable, in practice the energy needed to control the system may be disproportionately high to achieve control in practice.

Altogether, network control methods may represent an important avenue for the theoretical understanding of NF mechanisms and for the achievement of several possible important goals but further theoretical developments seem therefore necessary for control theory to realistically achieve clinical goals.

*2.2.2. Acting on brain function*
Much like standard stimulation techniques such as DBS or TMS, NF acts on some aspect of the brain's *dynamical* space, in order to navigate within the *functional* space, made observable by behaviour. How cognitive processes are identified and defined at the behavioural-cognitive level and in corresponding neurophysiological terms constitutes the critical point in the NF process (Micoulaud-Franchi et al., 2014b). While target states are *prima facie* framed in terms of brain dynamics, NF's effects are ultimately gauged in cognitive-behavioural terms. The fact that a given field affects some property of brain activity, e.g. some topological network property, does not entail, *per se*, a functional change *stricto sensu*. Such changes can be dynamically important but functionally neutral. Thus, what one needs to define is not dynamics *per se*, but its functional aspects.

But what is *functional* in brain activity? Answering this question is less straightforward than normally assumed, as this entails endowing the cognitive and dynamical spaces with their respective structure, characterizing a dynamics-to-function map, and defining the structure induced by this map. A smooth dynamics-to-behaviour map can sometimes be ensured, particularly when components and collective variables in the cognitive space can both be endowed with explicit differentiable analytical expressions (Kelso et al., 1998). However, in most contexts it is hard to conceive of the functional space as a smooth manifold, and describe it in terms of differential equations, so that the dynamics-to-function map is in general non-trivial. As a result, the structure of the functional space can only be revealed by projections onto an appropriate auxiliary space. Likewise, controlling *function*, rather than just dynamics, requires framing the target state in dynamical terms, defining functionally meaningful phase space partitions and a topology in such a space, and understanding *accessibility* of the corresponding dynamical regimes. How fine the dynamics-to-function map is in a given experimental setting determines the ability to address questions such as: how far is an observed behavioural state from some reference one? Can a given state be attained in the functional space, given the present one? What is the range of functionally neutral brain dynamical patterns?

Ultimately, characterizing the dynamics-to-function map implies not only defining what is *functional* in brain activity, but also addressing fundamental questions such as: what's the impact of cognitive demands on brain dynamics? How does behaviour result from dynamics? What aspects of brain activity can we use to define behaviour or, more ontologically, what aspects of brain activity does behaviour actually result from? Cognitive demands may for example be conceptualised as operators acting upon the symmetries of brain activity, and observed behaviour as a macroscopic property emerging from the renormalization of microscopic brain fluctuations and symmetry breaking of network connectivity (Papo, 2014; Pillai and Jirsa, 2017).

*2.3. Neurofeedback's mechanisms*

How is NF acting on the brain? Through what neural mechanisms does NF actually act? What neural structures and activities are summoned by NF? Such questions are of interest for all brain stimulations techniques, not only from a basic science viewpoint, but also from a technological one, as their answer is essential to things such as device development and optimization.

The mechanisms through which DBS exerts its effects, e.g. in the treatment for Parkinson's disease, have been investigated both computationally and experimentally (Wang et al., 2015; Santaniello et al., 2015; Saenger et al., 2017). While DBS's basic physics is known by construction, it has been suggested to improve motor symptoms by activating efferent fibres (Hashimoto et al., 2003), by modifying of oscillatory activity (Vitek, 2008), or by decoupling oscillations within the basal ganglia (Moran et al., 2012).

On the other hand, NF's physical processes and neural implementation mechanisms are far less understood (Sterman et al., 1996; Legenstein et al., 2010; Koralek et al., 2012; Birbaumer et al., 2013; Niv, 2013; Ros et al., 2014; Emmert et al., 2016; Davelaar, 2017). One important aspect examined in the literature is to do with the neural circuitry necessary for learning how to nudge brain dynamics in a functionally desirable way. It has been proposed that learning to self-regulate brain activity is akin to learning any other skill and that its underlying mechanisms is reinforcement learning (Birbaumer et al., 2013; Ros et al., 2014; Davelaar, 2017). Nevertheless, the following aspects are still poorly understood: how is neural dynamics modified as a result of NF? In particular, how does NF push brain dynamics to functionally desirable regions of the phase space? At an algorithmic level, there are various possible scenarios. For example, insofar as NF generally induces connectivity and possibly even topological changes, and that these changes may in turn be associated with dynamical events such as phase transitions (Kozma and Puljic, 2015), one possibility is that NF achieves its functional effects by forcing continuous transitions between symmetry groups or phases (Ros et al., 2014; Longo and Montévil, 2014; Papo, 2014). However, at an implementation level, the underlying neurophysiology through which NF is physically carried out is still largely unknown.

**3. Appraising neurofeedback**

Is NF *efficacious*? This fundamental questions has been addressed in various studies and meta-analyses (see e.g. Gevensleben et al., 2009, 2012; Kerson and Collaborative Neurofeedback Group, 2013; Niv, 2013; Canadian Agency for Drugs and Technologies in Health, 2014; Emmert et al., 2014; Sonuga-Barke et al., 2013; Vollebregt et al., 2014a,b; Micoulaud-Franchi et al., 2014a, 2016; Micoulaud-Franchi and Fovet, 2016; Schabus et al., 2017; Fovet et al., 2017; Alkoby et al., 2017; Sitaram et al., 2017; Thibault and Raz, 2017). But how is the success of a given intervention actually gauged?

Insofar are NF's effects are ultimately gauged in behavioural terms, a limiting element on the possible assessment of a given intervention is given by how well we can express the functional space and its impairment in terms of behaviour. This in turn is bounded by the strength of the topology of the *functional* space. For instance, evaluating efficacy of DBS on parkinsonian tremor is rather straightforward, as tremor can be gauged using the same dynamical variables as the corresponding brain oscillations. On the other hand, more cognitive symptoms e.g. in Parkinson's disease or in ADHD, let alone a symptom constellation, are not straightforward to map on a metric space and hard to convert into scalar target functions.

Another important question is that of knowing what criteria NF should satisfy for it to be deemed successful. While the dynamical characteristics of a successful NF protocol are of course to some extent goal- and time scale-dependent, an obvious evaluation criterion is represented by learning or persistence (Sherlin et al., 2011). But how can learning be framed in dynamical terms? At experimental time scales, learning may correspond to phase transitions in some order parameter describing collective brain activity either at system level or at some specified brain location. Learning may for instance be associated with a *bubbling transition*, a type of bifurcation resulting in a qualitative change in the way the attractor responds to dynamical perturbations or noise, whereby the system's typical behaviour may remain unchanged, while its attractor basin may be remodelled as a control parameter value crosses a critical value (Ashwin, 2006). At longer time scales, an important question is whether durable task-related change can be detected non-invasively in resting brain activity. Learning and learnability may for example be related to network properties of the system (Seung et al., 1992; Advani et al., 2013; Bassett and Mattar, 2017) or specific phases of brain activity (Carrasquilla, 2017). More generally, one may ask whether brain activity is more efficient in some dynamical sense as a result of successful NF (Ros et al., 2014), or more dynamically stable or robust. Studying these properties naturally leads to addressing issues such as the coupling between organizational architecture and dynamics, a hallmark of adaptive systems such as the brain.

Often, assessment of NF protocols understands the feasibility or goodness of NF in a given context taking for granted that the used target is the only possible or reasonable one, and overlooking possible alternatives to the used NF protocol. Only seldom are the following questions addressed: how good is a *target* for NF intervention? Is it a genuine control parameter for the dynamics? What can we do with the chosen target, i.e. could it be used to categorize different states or conditions, predict, stabilize, or target dynamical trajectories? While the NF target is implicitly thought of as a control parameter of some underlying dynamics, whether the distance from the current state (e.g. a dynamical regime or attractor) to a desired one is a simple and monotonic function of the control parameter is often largely unknown. If the chosen target is not a genuine control parameter for functional activity, distances along a scalar may not reflect the length along the path to the target. In the absence of such knowledge, a minimum requirement would seem that a target feature should fare well in a classification task between pre- and post-intervention conditions. However, how specific and sensitive these features are to the activity they are supposed to describe/control and how well they would fare as features in classification tasks is often poorly studied (Brandeis, 2011; Micoulad-Franchi et al., 2011). But, whether standard features would fare well or not in a classification task, the litmus test for the goodness of a target is given by the extent to which features can be used to *predict* dynamics, in some sense, or to model that system (Conant and Ashby, 1970). This would imply anything from extracting predictive information from brain signals (Still et al., 2010, 2012), to a complete read-out of the neural code (Panzeri et al., 2017).

## 4. Concluding remarks

Neuroscience is now arguably moving from a pure science to a technological stage, where devices and protocols are developed to interact directly with brain dynamics (rather than for instance its chemistry as in pharmacological intervention). But is our knowledge of the brain and of NF mechanisms sufficient to operate this transition?

The first obvious question is: how well shall we understand brain neurophysiology in order to optimize NF, i.e. on which mechanisms *should* NF protocols act in order to optimize behaviour? In the absence of a complete read-out of brain activity, there is a conceptual leap between simple features of brain dynamics and associated behaviour, an issue that becomes glaring in techniques such as invasive BCI, e.g. when predicting fine kinematic and kinetic parameters of limb movements to deliver to peripheral nervous system, or converting decoded signal to a language that the brain can understand. It is fair to assume that, while probably not necessary to NF's success (at least to some degree), understanding the dynamical mechanisms through which it actually works would help optimizing its potential. In fact, NF's efficacy may be proportional to how well it replicates neurophysiological closed feedback loop mechanisms.

So why do interventions work, at least to some extent, in the absence of complete or even satisfactory knowledge of the underlying mechanisms? A first answer to this question is of course related to the precision with which the functional outcome is evaluated. But another reason may be that, for a given coarse-graining level of the functional space, the underlying behaviour may in fact be described by *sloppy models*, i.e. multi-parameter models whose behaviour depends only on a few stiff combinations of parameters, with many sloppy parameter directions which have little or no impact on model predictions (Brown and Sethna, 2003). Using a partial combination of relevant parameters may therefore be enough to reach a region of the phase space functionally close to the desired range.

What information can we possibly extract non-invasively from ongoing activity that can be used as a feedback signal? What level of brain description should be chosen when devising NF protocols, i.e. when characterizing control parameters of brain activity? Theoretical results may suggest appropriate strategies to stabilize a given dynamics, e.g. via continuous feedback or by varying recurrent inhibitory connectivity (Hennequin et al., 2014). In addition, it would be important to define properties that can achieve the dynamical and functional criteria associated with successful NF, e.g. learning or functional robustness, or adaptability (Ma et al., 2009; Lan et al., 2012). Furthermore, one could reverse-engineer successful NF back to define the mechanisms through which it operates. This should help not only in describing at computational, algorithmic and implementation levels (Marr, 1982) why NF works, but also in understanding what brain description coarse-graining level optimizes NF targets. Network control strategies may for instance provide an indication as to the extent to which the brain works as a genuine network, as opposed to a set of connected modules (Papo et al., 2014).

Conversely, what sort of target can people learn to attain? For instance, while some evidence suggests that people can learn to modulate brain connectivity (Ruiz et al., 2014; Koush et al., 2017) or even particular transient spatially-extended patterns (Diaz Hernandez et al., 2016) there is as yet no clear indications as to the range of feasible NF targets, and questions such as whether or not the measure of brain activity needs to be unique ought to be addressed in earnest. One may conjecture that almost any target may be attainable, so long as it can be expressed in a convenient way, though ultimately the essential ingredient is that *cognitive control* be able to move brain dynamics to the desired functional regime. In other words, one thing is that a given control trajectory is theoretically feasible, and another one is that a given subject proves able to push the current state to the desired one. Even when targets can be expressed in scalar form, how exactly, *viz.* through what neural mechanisms these targets are approached and achieved, is still poorly understood.

Finally, given NF's non-invasiveness, an intriguing issue is whether elements of NF can help improving invasive brain stimulation techniques and vice versa. For instance, DBS, which constitutes a therapeutic option in various pathologies, is a chronic stimulation technique, but is in general not equipped with a feedback control loop. On the other hand, NF is a time-limited stimulation technique with a control loop. Its effects are based on change in neurophysiological parameters, though exactly what aspects of neural activity and how durable these effects are is still a matter of debate. Device-mediated stimulation techniques will necessarily need to cope with issues of goal attainability, given the physical limits of available technology (Marblestone et al., 2013), and should all benefit from the addition of *feedback loops*. On the other hand, a better understanding of NF mechanisms may supply some answers to the main limits to BCI development, i.e. achieving safe, accurate and robust access to brain signals, and may help equipping chronic stimulation techniques with effective closed-loop strategies and possibly even supersede them, in some instances.

In conclusion, identifying general goals and targets of neural stimulation and, in particular, of NF, entails questioning not only how people learn to steer their own activity in a goal-directed way, but also how we describe (functional) brain activity in general. While the ability to perturb behaviour using brain dynamics will provide an important tool to reveal functional mechanisms underlying changes in behaviour, improving NF will require better understanding of various fundamental aspects of brain dynamics, and a more explicit definition of what is functional in brain activity and of brain-behaviour relationships.

**Acknowledgements**

The author acknowledges financial support from the program *Accueil de Talents* of the Métropole Européenne de Lille.